\documentstyle[aps,multicol,epsf]{revtex}
\begin{document}
%%%%%%%%%%%%%%%%%%%%%%%%%%%%%%%%%%%%%%%%%%%%%%%%%%%%%%%%%%%%%%%%%%%%%%%%
\draft
\title{Spinodal-assisted crystallization in polymer melts}
\author{P.~D. Olmsted,$\!^{\ast}$ W.~C.~K. Poon,$\!^{\dagger}$
T. C. B. McLeish,$\!^{\ast}$
N. J.Terrill$\!^{\ddagger}$ and A. J. Ryan$^{\ddagger}$}
\address{$^{\ast}$Department of Physics \& Astronomy and Polymer IRC,
University of Leeds, Leeds, LS2 9JT, England {\tt
(p.d.olmsted@leeds.ac.uk)}; $^{\dagger}$Department
of Physics \& Astronomy, The University of Edinburgh, Mayfield Road,
Edinburgh, EH9 3JZ, Scotland {\tt (w.poon@ed.ac.uk)}; 
$^{\ddagger}$Department of Chemistry,
University of Sheffield, Brookhill, Sheffield, S3 7HF, England}
\maketitle
\makeatother
%%%%%%%%%%%%%%%%%%%%%%%%%%%%%%%%%%%%%%%%%%%%%%%%%%%%%%%%%%%%%%%%%%%%%%%%
\begin{abstract}
  Recent experiments in some polymer melts quenched below the melting
  temperature have reported spinodal kinetics in small-angle X ray
  scattering before the emergence of crystalline structure. To
  explain these observations we propose that the coupling between
  density and chain conformation induces a liquid-liquid binodal
  within the equilibrium liquid--crystalline solid coexistence region.
  A simple phenomenological theory is developed to illustrate this
  idea, and several experimentally testable consequences are
  discussed.  Shear is shown to enhance the kinetic role of the hidden
  binodal.
%%%%%%%%%%%%%%%%%%%%%%%%%%%%%%%%%%%%%%%%%%%%%%%%%%%%%%%%%%%%%%%%%%%%%%%%
%PACS: 
\end{abstract}
%%%%%%%%%%%%%%%%%%%%%%%%%%%%%%%%%%%%%%%%%%%%%%%%%%%%%%%%%%%%%%%%%%%%%%%%
\begin{multicols}{2}
\narrowtext
%%%%%%%%%%%%%%%%%%%%%%%%%%%%%%%%%%%%%%%%%%%%%%%%%%%%%%%%%%%%%%%%%%%%%%%%

Upon cooling a crystallizable polymer melt a hierarchy of ordered
structures emerges \cite{strobl}. First there are crystalline
`lamellae', comprising regularly packed polymer chains, each of which
is ordered into a specific helical conformation. These lamellae
interleave with amorphous layers to form `sheaves', which in turn
organise to form superstructures (\textit{e.g.} spherulites). These
structures may be probed by various techniques: {\it e.g.\/}
wide-angle X ray scattering (WAXS) is sensitive to atomic order within
lamellae (`Bragg peaks'), while small-angle X ray scattering (SAXS)
probes lamellae and their stacking.  Inspired by recent experiments,
we propose a model for the earliest stages of ordering in a
supercooled polymer melt, and discuss several experimentally-testable
consequences, including strain and pressure effects.

 In a supercooled simple liquid, the following picture
\cite{kinetics} is widely accepted. Nuclei of the lower free energy
(crystal) phase are constantly formed by thermal fluctuations. But the
cost of creating an interface means that only large enough nuclei can
grow --- the melt is {\it metastable}. An induction time, $\tau_{i}$,
elapses before the probability of forming such `critical nuclei'
becomes significant. This picture is usually deemed appropriate for
polymer melts; instead effort is focused on explaining the anisotropic
shape and growth rate of crystal nuclei \cite{gerhard}.

In the `classical' picture of polymer melt crystallization we expect
and observe Bragg peaks in WAXS {\it after} an induction period
$\tau_i$.  SAXS accompanies the WAXS, corresponding to interleaved
crystal lamellae and amorphous regions \cite{strobl}.  No SAXS is
expected during $\tau_{i}$. However, recent experiments have reported
SAXS peaks during the induction period and {\it before} the emergence
of Bragg peaks. Initially the SAXS peak intensity grows exponentially,
and it may be accurately fitted to Cahn-Hilliard (CH) theory for {\it
  spinodal decomposition} --- the spontaneous growth of fluctuations
indicative of thermodynamic {\it instability} \cite{gunton}.  The peak
moves to smaller angles in time, stopping when Bragg peaks emerge. By
fitting to CH theory an extrapolated spinodal temperature (at which
the melt first becomes unstable towards local density fluctuations)
can be obtained. Spinodal kinetics have been reported in different
polymer melts: poly(ethylene terphthalate) (PET) \cite{imai},
poly(ether ketone ketone) (PEKK) \cite{hsiao}, polyethylene (PE) and
isotactic polypropylene (i-PP) \cite{ryan,foot2}. Despite these recent
results, no coherent model exists for these phenomena.

Such observations can be explained by appealing to the concept of a
`metastable phase boundary', a common strategy in metallurgy
\cite{cahn,microstruct}.  Consider an alloy quenched into a region of
its phase diagram where we expect coexistence between two phases, say,
$\alpha + \beta$. If the two phases are symmetry-unrelated there is
always an energy barrier for phase separation, and we do {\it not}
expect to see spinodal (unstable) dynamics. Nevertheless, spinodal
dynamics and textures have been observed in some such cases. Some time
ago, Cahn suggested \cite{cahn} that this could be due to a metastable
phase boundary for coexistence between two symmetry-related phases
(say, $\alpha + \alpha^{\prime}$) {\it buried within} the equilibrium
$\alpha + \beta$ coexistence region. Recently, similar ideas are used
to explain slow kinetics in the formation of colloidal crystals in
colloid-polymer mixtures \cite{faraday} and globular protein
crystallization \cite{protein}.

\begin{figure} \displaywidth\columnwidth \epsfxsize=3.0truein
\centerline{\epsfbox[90 90 700 490]{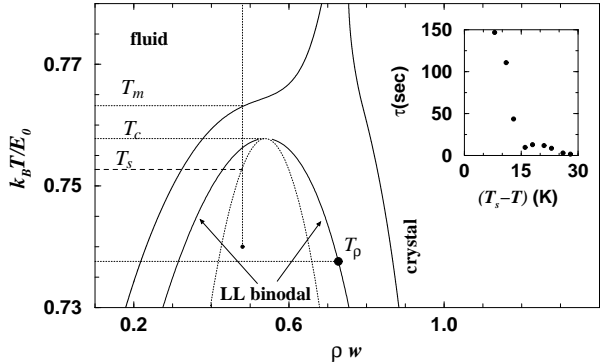}} 
\caption{Proposed
generic phase diagram for a polymer melt calculated as described in
the text. $T_{m}$ and $T_{s}$ are the melting and spinodal
temperatures encountered along the ({constant density}) quench path
(dotted line). Parameters used are $R M_b=0.8, k_{\scriptscriptstyle
B}T_{\ast}= 0.29\, E_0, v=1.4 \,E_0 w,$ $\lambda = 0.1\, a v_0,$
$b = -0.4\, (v_0a^3/E_0)^{1/2}$, $c=0.5\, a^2 v_0/E_0,$ and
$\alpha=0.8\,/w$.  Inset shows the measured induction time as a
function of temperature for isotactic polypropylene \protect\cite{ryan}.}
\label{fig:phdiag}
\end{figure}

Similarly, a plausible explanation for the observation of spinodal
dynamics in polymer melts is the presence of a metastable liquid-liquid
(LL) phase coexistence curve (or `binodal') buried inside the
equilibrium liquid-crystal coexistence region, Fig.~\ref{fig:phdiag}.
Quenching sufficiently below the equilibrium melting point $T_{m}$, we
may cross the spinodal associated with the buried LL binodal at
temperature $T_{s} < T_{m}$. Below, we give a physical mechanism that
can give rise to such a metastable binodal, calculate the phase diagram
using a phenomenological free energy, and delineate some consequences
of our model.

Our starting point is the unremarkable statement that, in order to
crystallize, polymer chains must adopt the correct {\it conformation}.
For example, chains in crystalline PE have the all-{\it trans} (or
`zig-zag') conformation, while in the melt the conformation is
randomly {\it trans} or {\it gauche}. Generally, the preferred
conformation is some form of helix. Furthermore, the radius of
gyration of a (very long) chain changes little during crystallization,
suggesting \cite{fischer} that neighboring segments adopt the correct
conformation and crystallize `in situ'. It is commonly assumed that
conformational (intrachain) and crystalline (interchain) ordering
occur {\it simultaneously}. We suggest, in light of recent
experiments, that these processes can occur {\it sequentially}. To
motivate this suggestion we examine more closely the physics of
conformation changes.

%%%%%%%%%%%%%%%%%%%%%%%%%%%%%%%%%%%%%%%%%%%%%%%%%%%%%%%%%%%%%%%%%%%%%%%%
%%%%%%%%%%%%%%%%%%%%%%%%%%%%%%%%%%%%%%%%%%%%%%%%%%%%%%%%%%%%%%%%%%%%%%%%

In a melt, it is believed that chain conformation alone cannot drive a
phase transition. However, conformation is coupled to density. Chains
with the `correct' (helical) conformation typically pack more densely
than those with more or less random conformations. Moreover, the
energy barriers between different rotational isomeric states (RIS) are
density-dependent \cite{chandler}. We now show that {\it
  conformation-density coupling can induce a LL phase transition.} A
phenomenological free energy which incorporates these effects is a
function of the following order parameters: the (average) mass density
$\bar{\rho}$; the coefficients $\{\rho_{\bbox{q}}\}$ in the Fourier
expansion of the crystal density in terms of the appropriate stars of
reciprocal lattice vectors $\{\bbox{q}\}$ (essentially the intensities
of Bragg peaks) \cite{landaustat1}; and the occupancies $\{\eta_{i}\}$
of various RIS (and thus chain conformation). To illustrate the
principles, we assume that a single $\rho_{\bbox{q}} \equiv
\rho_{\ast}$ and a single $\eta$ suffice, corresponding to a
fictitious polymer with body-centered cubic crystal structure
\cite{bcc} and two RIS.  The free energy per unit volume has three components:
\begin{equation} 
f = f_{0}(\bar{\rho}) + f_{\ast}(\bar{\rho},\rho_{\ast}) +
f_{\eta}(\eta,\bar{\rho}, \rho_{\ast}). \label{eq:freengy}
\end{equation} 
The equation of state $\bar{\rho}\partial
f/\partial\bar{\rho}\!-\!f=\!p$ determines the $T\!-\!\bar{\rho}$
isobar at pressure $p$.  The first term, $f_{0}$, is the free energy
of a melt with random chain conformations. Equation-of-state fits to
polymer liquids suggest the following form
\begin{equation}
f_{0}({\bar{\rho}}) = 
R\, k_{\scriptscriptstyle B}T \bar{\rho}\ln\left[(1/\bar{\rho})-w\right],
\label{eq:f0}
\end{equation}
where $R$ and $w$ are widely tabulated
\cite{polymerhandbook}.  The (bare) Landau free energy of
crystallization is taken to be \cite{landaustat1,bcc}
\begin{equation}
f_{\ast}(\bar{\rho},\rho_{\ast}) = \bar{\rho} \left[\case12
a\left(\bar{\rho},T\right) \rho_{\ast}^2 + \case13 b\,\rho_{\ast}^3 +
\case14 c\,\rho_{\ast}^4\right],
\label{eq:freec}
\end{equation}
For simplicity we let $a(\bar{\rho},T)\equiv a_0 k_{\scriptscriptstyle
  B} \left[T - T_{\ast} \left(1 + \alpha \bar{\rho}\right)\right]$,
where $\alpha$ and $T_{\ast}$ account for the enhancement of
crystallization due to increased density. $f_0 + f_{\ast}$ has a
double well structure and gives a (bare) first-order transition
between amorphous ($\rho_{\ast} = 0$) and crystalline ($\rho_{\ast}
\neq 0$) states.

$f_{\eta}$ describes how the distribution of chain conformations
varies smoothly from random ($\eta = 0$) to totally ordered (helix,
$\eta = 1$) as the temperature is lowered to zero \cite{flory}.  In
isolation a polymer thermally populates its RIS with a Boltzmann
distribution $P_{\alpha}\sim \exp\{-\beta E_{\alpha}\}$, where
$E_{\alpha}$ is the energy of state $\phi_{\alpha}$ relative to the
ground state $\phi_0$, and $\beta=1/k_{\scriptscriptstyle B}T$. As the
temperature is lowered the mean occupancy $\eta$ of state $\phi_0$,
relative to the $T=\infty$ disordered state ($\eta=0$), increases. We
describe this process, for a two-state model, by
\begin{equation} f_{\eta}({\eta}, \bar{\rho},\rho_{\ast})
= {k_{\scriptscriptstyle B}T 
\bar{\rho}\over 2 M_b}\left[\eta^2 \cosh^2({\beta E\over 2})- \eta\sinh(\beta
E)\right],
\label{eq:feta}
\end{equation} 
where $M_b$ is the mass of a monomer with characteristic volume $v_0=w
M_b$.  Minimizing $f_{\eta}$ over $\eta$ yields the correct Boltzmann
distribution $\eta(T)=\tanh(\beta E/2)$. We choose \cite{eta}
\begin{equation}
E(\bar{\rho},\rho_{\ast}) = E_0 + v \bar{\rho} + \lambda
\rho_{\ast}^2. \label{eq:ris}
\end{equation}
As more bond sequences occupy the ground state, monomers can rearrange
to pack tighter and reduce the excluded volume interaction (hence the
perturbation $v\bar{\rho}$). A positive $v$ encourages phase
separation to take advantage of this density-conformation coupling.
Similarly, adjacent ground state sequences enhance crystallization
(hence the term $\lambda \rho_{\ast}^2)$. The $\lambda$ term is
quadratic in $\rho_{\ast}$ by symmetry \cite{bcc}.

\begin{figure}
\displaywidth\columnwidth
\epsfxsize=3.0truein
\centerline{\epsfbox[0 0 718 469]{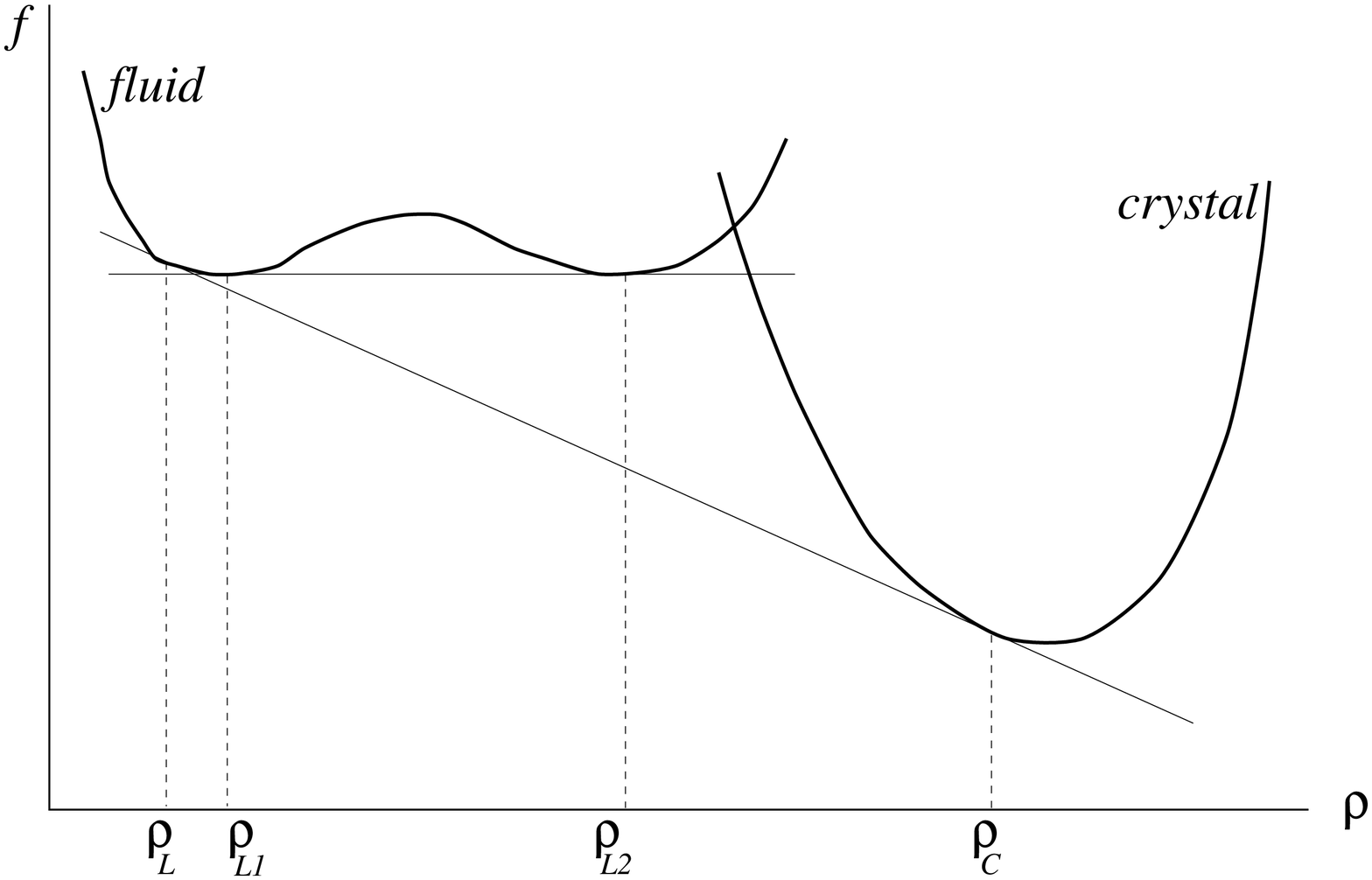}}
\caption{Schematic free energy density plots at a particular
  temperature. At this temperature, a melt with density $\rho_{L} <
  \rho < \rho_{c}$ will phase separate into coexisting liquid and
  crystal phases at densities $\rho_{L}$ and $\rho_{c}$. The common
  tangent drawn on the liquid branch alone gives the densities,
  $\rho_{L1}$ and $\rho_{L2}$, on the metastable LL binodal, for which
  see Fig.~\protect\ref{fig:phdiag}.}
\label{fig:tangent} 
\end{figure}

\begin{figure} 
\displaywidth\columnwidth
\epsfxsize=3.0truein
\centerline{\epsfbox[0 0 558 503]{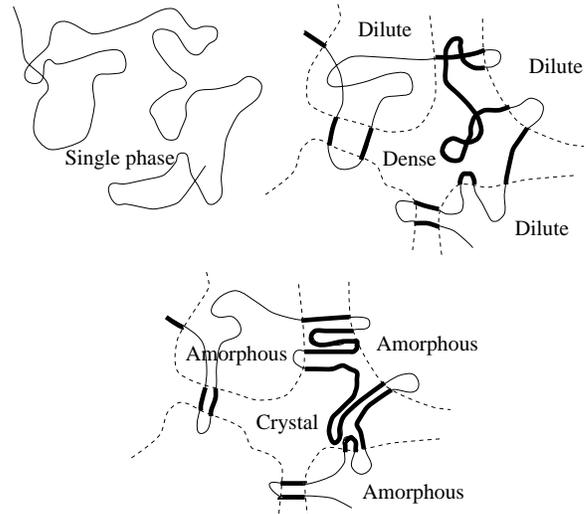}}
\caption{Schematic representation of the late-stage
  spinodal texture for coexisting liquid phases with different
  conformations, showing a single chain; thin line = disordered
  conformation, thick line = correct (helical) conformation for
  crystallization. Each chain is a `conformational copolymer'.}
\label{fig:coex}
\end{figure}

To calculate the phase diagram in the temperature-densiy
($T-\bar{\rho}$) plane we first minimize $f$ with respect to $\eta$.
Note that minimizing $f_{\eta}$ over $\eta$ and expanding $f$ in
$\bar{\rho}$ renormalizes $a(\bar{\rho},T)$ and $c$ in
Eq.~(\ref{eq:freec}) and, at sufficiently small $T$, destabilizes the
homogeneous melt.  Physically, the system gives up conformational
entropy to relieve packing frustration, and separates into a dense,
more ordered liquid and a less dense and less ordered liquid.  The
renormalization of $c$ lowers the barrier to crystallization in the
dense, high-$\eta$ liquid.  Next, $f$ is minimized with respect to
$\rho_{\ast}$ to give a final free energy with two branches,
$\hat{f}(\bar{\rho}, \rho_{\ast} = 0)$ (liquid) and
$\hat{f}(\bar{\rho}, \rho_{\ast} \neq 0)$ (solid). The common tangent
constructed between these two branches at any temperature gives the
densities of coexisting liquid and crystal phases at that temperature,
Fig.~\ref{fig:tangent} \cite{dehoff}. At low enough temperatures, the
liquid branch gives a spinodal buried entirely within the equilibrium
liquid-crystal coexistence region given as usual by the points of
inflection, $\partial^{2}\hat{f}/\partial \bar{\rho}^{2} = 0$. If a
melt is quenched inside this spinodal, it will phase separate into two
coexisting liquids, given by the common tangent construction, with a
coarsening interconnected domain texture, Fig.~\ref{fig:coex}. The two
liquids differ in their distributions of conformations, with the
denser liquid adopting a distribution closer to that needed for
crystalline packing.

%%%%%%%%%%%%%%%%%%%%%%%%%%%%%%%%%%%%%%%%%%%%%%%%%%%%%%%%%%%%%%%%%%%%%%%%%%%%%
%%%%%%%%%%%%%%%%%%%%%%%%%%%%%%%%%%%%%%%%%%%%%%%%%%%%%%%%%%%%%%%%%%%%%%%%%%%%%

To calculate Fig.~\ref{fig:phdiag} we fix the dimensionless parameter
$R M_b = 0.8$, appropriate for PE \cite{polymerhandbook}, leaving
$v/(E_0 w)$ as the only adjustable parameter to determine the spinodal
temperature $T_s$. For $v=1.4\,E_0 w$ we find an LL critical point at
$k_{\scriptscriptstyle B}T_c=0.75 E_0$ and $\rho_c w = 0.53$.
Experiments on PE, for which $\rho w \simeq 0.685$ ($w=0.875\, {\rm
  cm}^3\,{\rm g}^{-1}$ \cite{polymerhandbook} and $\rho=0.783\, {\rm
  g\, cm}^{-3}$ \cite{mccoy}), give $k_{\scriptscriptstyle B}T_s
\simeq 0.86\,E_0$ \cite{ryan} (using $E_0=930\,{\rm
  cal/mole}$\cite{chandler}). The value for $v$ corresponds to a
relative change of $E$ by of order $+0.58$ on going from single chain
to melt conditions, Eq.~\ref{eq:ris}. This agrees, in sign and
magnitude, with the known behaviour of common polymers \cite{mccoy}.
The agreement over {\it sign} is particularly significant. We suggest
that density and conformation work cooperatively, \textit{i.e.} $v >
0$; no LL binodal within the required temperature range was obtained
for $v <0$. Even a crude Landau theory, therefore, puts constraints on
allowable physical mechanisms. Our choice of crystallization
coefficients ($\alpha, b, c, \lambda$) (caption,
Fig.~\ref{fig:phdiag}), gives a reasonable value for $T_{m}$, but the
crudeness of Landau theory for first order transitions renders
detailed fitting somewhat meaningless.

One of the coexisting liquids is closer in density and conformation to
the crystal phase than the original melt, and has a lower energy
barrier, $\Delta(T,\bar{\rho})$, to crystallization, so inducing
`spinodal-assisted nucleation'. We expect $\Delta(T,\bar{\rho})$ to
decrease with increasing quench depth below $T_{s}$ (and hence widening
LL coexistence gap).  Indeed, in our simplified model, we find
$\Delta(T,\bar{\rho}) = 0$ at temperature $T_{\rho} < T_{s}$. The
induction time, $\tau_{i}$, is then
\begin{equation} \tau_{i} \sim
\tau_{s} + \mbox{const.} \times e^{\Delta(T,\bar{\rho})/k_{B}T}\;,
\end{equation} 
where $\tau_{s}$ is the time to reach an intermediate spinodal
texture, Fig.~\ref{fig:coex}, in which the regions have (almost) the
coexisting LL densities. We expect $\tau_{s}$ to be weakly (at most
power law) dependent on temperature. The exponential term accounts for
the barrier to nucleate a crystal from the dense liquid.  The strong
temperature dependence of $\tau_{i}$ should change over to a much
weaker dependence at some $T_{\rho} < T_{s}$, where
$\Delta(T_{\rho},\bar{\rho}) \alt k_{\scriptscriptstyle B}T_{\rho}$.
This has been found in i-PP (inset, Fig.~\ref{fig:phdiag})
\cite{ryan}. 

The characteristic length scale associated with the developing spinodal
texture gives rise to a SAXS peak, which evolves initially according to
CH theory \cite{kawasaki}. The coarsening of this texture is observed
to be arrested at the end of the induction period (typical scale
$\xi_{m}$), when Bragg peaks appear in WAXS \cite{ryan}.  It is at
present unclear how the spinodal texture at the end of the induction
period evolves into spherulites. However, the final spinodal texture
length scale $\xi_{m}$ evidently controls the thickness of the first
crystal lamellae. Moreover, large stress will develop once one of the
two liquids in a bicontinuous texture, Fig.~\ref{fig:coex},
crystallizes. We expect such a texture to fragment into individual
crystalline lamellae.
%%%%%%%%%%%%%%%%%%%%%%%%%%%%%%%%%%%%%%%%%%%%%%%%%%%%%%%%%%%%%%%%%%%%%%%%
%%%%%%%%%%%%%%%%%%%%%%%%%%%%%%%%%%%%%%%%%%%%%%%%%%%%%%%%%%%%%%%%%%%%%%%%

Our arguments so far have been based on conformation-density coupling.
Once a polymer segment has adopted the correct (helical) conformation
its persistence length should increase, which couples to the {\it
  orientational} order of chains. Indeed, depolarized light scattering
by Imai and coworkers has suggested the existence of orientational
fluctuations during the spinodal phase of a crystallizing PET melt
\cite{imai2}. Provided that orientational ordering is not strong
enough to induce a separate transition, then the inclusion of a
nematic order parameter in Eq.~\ref{eq:freengy} only renormalises the
coefficients in $\eta$-dependent terms. In some cases, the increasing
chain stiffness accompanying conformational order may be sufficient to
drive an isotropic $\rightarrow$ nematic transition, resulting in a
{\it three}-step process: melt $\rightarrow$ (isotropic) liquid(1) +
liquid(2), followed by liquid(2) $\rightarrow$ nematic $\rightarrow$
crystal. This possibility should be investigated. It should also be
possible to model specific polymers, {\sl e.g.\/} using density
functional theory \cite{mccoy}, augmented to include the effects
leading to Eq.~\ref{eq:ris} \cite{chandler}.

Until recently, spinodal scattering was mainly observed in polymer
melts crystallizing {\it under shear} \cite{strobl,foot2,miller}.
This may be understood in a natural way within the present framework.
Shear (and extensional) flow couples principally to the orientation of
polymer segments, hence straightening chains and enhancing $\eta$,
thereby biasing the tendency towards LL separation.  A simple way to
incorporate this is to renormalise the activation energy $E$ as
$E-v_0\sigma$ where $\sigma$ is the stress.  It is highly suggestive
that, for appropriate values of stress under strong flow (the plateau
modulus $G_0$) and volume ($v_0$ above), the LL binodal of Fig.~1 is
shifted upward significantly (by $\delta T_s \sim 0.01
E_0/k_{\scriptscriptstyle B}$). Flow will shift the liquid-solid
coexistence curve much less because the regions with crystalline order
will resist deformation.

Our simple theory suggests several interesting experiments. First, and
most directly, conformational fluctuations just above $T_{s}$ could be
detected and studied, \textit{e.g.}, by Raman spectroscopy \cite{strobl},
perhaps simultaneously with depolarised light scattering (to monitor
orientational fluctuations). Secondly, on approaching a spinodal line,
various properties (\textit{e.g.} correlation length) should exhibit
power-law divergences. The observation of such divergences will lend
much weight to the correctness of our model.  Thirdly, the point at which
the LL spinodal is encountered in a quench can be modified by pressure,
Fig.~\ref{fig:phdiag}. In particular, it may be possible to access the LL
critical point, $T_{c}$: recent simulations suggest a massive enhancement
of the nucleation rate in the vicinity of $T_{c}$ \cite{frenkel}. More
generally, the coupling of density to (molecular) structural order
parameters is an emerging generic theme in the study of supercooled
liquids (water $\rightarrow$ amorphous ice, \cite{harrington97}; polymer
melts near the glass transition \cite{kanaya94}).  Finally, processes
such as surface nucleation could give rise to a SAXS peak, but are
unlikely to follow Cahn-Hilliard kinetics. Also, effects other than
conformation and orientation ({\it e.g.\/} polydispersity) may induce
LL phase separation.  Experiments on monodisperse alkanes are under way
to address this possibility.

{\it Acknowledgements} WCKP was funded by the Nuffield Foundation, and 
NJT and AJR by the UK EPSRC (GR/K33767).

%%%%%%%%%%%%%%%%%%%%%%%%%%%%%%%%%%%%%%%%%%%%%%%%%%%%%%%%%%%%%%%%%%%%%%%%

%%%%%%%%%%%%%%%%%%%%%%%%%%%%%%%%%%%%%%%%%%%%%%%%%%%%%%%%%%%%%%%%%%%%%%%%
\end{multicols}
%%%%%%%%%%%%%%%%%%%%%%%%%%%%%%%%%%%%%%%%%%%%%%%%%%%%%%%%%%%%%%%%%%%%%%%%

\begin{references}

\bibitem{strobl} G. Strobl, {\it The Physics of Polymers}, Springer,
Berlin (1996).

\bibitem{kinetics} J. Frenkel, {\it The Kinetic Theory of Liquids}, Oxford
(1946). 

\bibitem{gerhard} G. Goldbeck-Wood, in J.~P. van der Eerden and O.~S.~L.
Bruinsma, Eds., {\it Science and Technology of Crystal Growth}, Kluwer,
Dordrecht (1995), 313. 

\bibitem{gunton} J.~D.~Gunton, M.~San Miguel and P.~S.~Sahni, in {\it
    Phase Transitions and Critical Phenomena\/}, edited by C.~Domb and
  M.~S.  Green (Academic, New York, 1983), Vol.~8.

\bibitem{imai} M. Imai, K. Kaji and T. Kanaya, Macromol. {\bf 27}, 7103
(1994). 

\bibitem{hsiao} T.~A. Ezquerra, E. Lopezcabarcos, B.~S. Hsiao, and
F.~J. Baltacalleja, Phys. Rev.~\textbf{E 54}, 989 (1996).

\bibitem{ryan} Polypropylene, N.~J. Terrill, J.~P.~A. Fairclough, B.~U.
Komanschek, R.~J. Young, E. Towns-Andrews and A.~J. Ryan, Polymer {\bf
    39}, 2381 (1998); polyethylene; N. J. Terrill and A.~J. Ryan,
  unpublished.

\bibitem{foot2} Spinodal kinetics are also observed in polymer melts
  crystallizing under shear. For poly(vinylidene Fluoride) (PVF), see M.
  Cakmak, A. Teitge, H.~G. Zachmann, and J.~L. White, J. Polymer
  Sci: Pt. B Polymer Phys.  {\bf 31}, 371 (1993); for PET, see R.
  G\"{u}nther, Ph.D.  thesis, Universit\"{a}t Mainz (1994), data quoted
  by Strobl, ref.  \cite{strobl}, section 4.2.2.

\bibitem{cahn} J.~W. Cahn, Trans. Metall. Soc. AIME {\bf 242}, 166
(1968). 

\bibitem{microstruct} See, \textit{e.g.}, J.~W. Martin, R.~D. Doherty and B.
  Cantor, {\it Stability of microstructures in metallic systems},
  Cambridge University Press, Cambridge (1997), section 3.2.
  
\bibitem{faraday} W.~C.~K. Poon, A.~D. Pirie and P.~N. Pusey, Faraday
  Discuss. {\bf 101}, 65 (1995); M. R. L. Evans, W. C. K. Poon and M.
  E.  Cates, Europhys. Lett. \textbf{38}, 595 (1997).

\bibitem{protein} W.~C.~K. Poon, Phys. Rev. E {\bf 55}, 3762 (1997).

\bibitem{fischer} M. Dettenmaier, E.~W. Fischer and M. Stamm, Colloid
Polymer Sci. {\bf 258}, 343 (1980).

\bibitem{chandler} L.~R. Pratt, C.~S. Hsu, and D. Chandler, 
J. Chem. Phys. {\bf 68}, 4202 (1978).

\bibitem{landaustat1} L.~D. Landau and E.~M. Lifschitz, {\it Statistical
Physics, 3rd Edition, Part 1\/}, Pergamon, Oxford (1980).

\bibitem{bcc} S. Alexander and J. McTague, Phys. Rev. Lett. {\bf 41},
702 (1978).

\bibitem{polymerhandbook} J. Brandrup and E.~H. Immergut, {\it Polymer
Handbook}, 3rd ed. (Wiley, New York, 1989).

\bibitem{flory} P.~J. Flory, {\it Statistical Mechanics of Chain
Molecules}, Oxford University Press, New York (1989).

\bibitem{eta} The coupling $E(\bar{\rho},\rho_{\ast})$ accounts for
  the effects of density and crystallinity on the RIS energies. We can
  also include free energy terms such as $- \eta \bar{\rho}^2$ and $ -
  \eta\bar{\rho} \rho_{\ast}^2$, which have, in principle, other
  physical justifications (excluded volume).  Such terms emerge
  naturally upon expanding Eq.~\ref{eq:feta} in the order parameters,
  but a description in terms of the full $E(\bar{\rho},\rho_{\ast})$
  assures that $\eta\le 1$ and gives the necessary qualitative form
  for the free energy.
  
\bibitem{dehoff} T. Dehoff, {\it Thermodynamics in Materials Science},
  McGraw Hill, New York (1992).
  
\bibitem{kawasaki} Fluid-fluid phase separation near a critical point
  obeys `model H' dynamics, which is equivalent to CH theory; see
  K.~Kawasaki, in {\it Phase Transitions and Critical Phenomena\/},
  edited by C.~Domb and M.~S.  Green (Academic, New York, 1976),
  Vol.~5a.

\bibitem{imai2} M. Imai, K. Kaji, T. Kanaya, and Y. Saka, Phys. Rev. B
{\bf 52}, 12696 (1995). 

\bibitem{mccoy} J.~D. McCoy, K.~G. Honnell, K.~S. Schweizer, and J.~G.
Curro, J. Chem. Phys. {\bf 95}, 9348 (1991).

\bibitem{miller} R.~L. Miller (Ed.), \textit{Flow-induced
    Crystallization in Polymers Systems}, (Gordon and Breach, New
  York, 1979).

\bibitem{frenkel} P.~R. ten Wolde and D. Frenkel, Science {\bf 277},
1975 (1997). 

\bibitem{harrington97}
S. Harrington, R. Zhang, P.~H. Poole, F. Sciortino, and H.~E.
Stanley, Phys. Rev. Lett. {\bf 78},  2409  (1997).

\bibitem{kanaya94}
T. Kanaya, A. Patkowski, E.~W. Fischer, J. Seils, H. Glaser,
and K. Kaji, Acta Poly. {\bf 45},  137  (1994).


\end{references}
\end{document}